\documentclass[twocolumn,preprintnumbers]{revtex4}%
\usepackage{amsmath}
\usepackage{amssymb}
\usepackage{graphicx}
\usepackage{dcolumn}
\usepackage{bm}
\usepackage{amsfonts}%
\begin{document}
\title{Upper Bound Imposed upon Responsivity of Optical Modulators}
\author{Eyal Buks}
\email{eyal@ee.technion.ac.il}
\affiliation{Department of Electrical Engineering, Technion, Haifa 32000, Israel }
\date{\today }

\begin{abstract}
We study theoretically the responsivity of optical modulators. \ For the case
of linear response we find using perturbation theory an upper bound imposed
upon the responsivity. \ For the case of two mode modulator we find a lower
bound imposed upon the optical path required for achieving full modulation
when the maximum birefringence strength is given.

\textit{OCIS} codes: 060.4080, 230.4110, 350.1370

\end{abstract}
\maketitle





\textit{Introduction} - Optical modulators are devices of great importance for
telecom and other fields. \ These devices allow controlling the transmission
$T$ ($0\leq T\leq1$) between input and output ports by applying some external
perturbation. \ One of the key characterization of optical modulators is their
responsivity, namely the dependence of $T$ on the applied external
perturbation. \ Enhancing the responsivity is highly desirable for many
applications. \ This raises the question what is the largest possible
responsivity that can be achieved for a given perturbation mechanism employed.
\ Here we consider this question for the case of linear modulators. \ We show
that the linearity of such devices imposes upper bound on the responsivity.

\textit{Perturbation Theory - }Consider an optical modulator consisting of an
optical path of length $\Delta s=s_{1}-s_{0}$. \ Here we consider the case
where the light passes the optical path only once (contrary to the case of a
resonator where multiple reflections occur). \ At each point $s$ along the
optical path the field is expanded using some local orthonormal basis. \ Using
the Dirac notation (\textit{bra} and \textit{ket}) \cite{Weissbluth} the field
at point $s$ is denoted as $\left\vert \psi\left(  s\right)  \right\rangle ,$
which represents a column vector of amplitudes. \ The equation of motion is
given by%
\begin{equation}
\frac{d}{ds}\left\vert \psi\right\rangle =i\mathcal{K}\left\vert
\psi\right\rangle ,
\end{equation}
where the Hermitian linear operator $\mathcal{K}$\ is the \textit{Hamiltonian}
of the system. \ Consider the effect of adding a small perturbation
$\mathcal{\varepsilon K}_{1}\left(  s\right)  $ to the unperturbed Hamiltonian
$\mathcal{K}_{0}$, namely%

\begin{equation}
\mathcal{K}\left(  s\right)  \mathcal{=K}_{0}\left(  s\right)
\mathcal{+\varepsilon K}_{1}\left(  s\right)  ,
\end{equation}
where $\left\vert \varepsilon\right\vert <<1$ is a small real parameter. \ For
any given $\varepsilon$ the final state $\left\vert \psi_{f}\right\rangle
=\left\vert \psi\left(  s_{1}\right)  \right\rangle $ is related to the
initial state $\left\vert \psi_{i}\right\rangle =\left\vert \psi\left(
s_{0}\right)  \right\rangle $ by the relation%

\begin{equation}
\left\vert \psi_{f}\right\rangle =U\left(  \varepsilon\right)  \left\vert
\psi_{i}\right\rangle ,
\end{equation}
where $U\left(  \varepsilon\right)  $ is the $s$ evolution operator for the
Hamiltonian $\mathcal{K=K}_{0}\mathcal{+\varepsilon K}_{1}$ from $s=s_{0}$ to
$s=s_{1}$. \ The final state $\left\vert \psi_{f}\right\rangle $ is filtered
by a polarizer having a normalized state $\left\vert \psi_{p}\right\rangle $.
\ The transmission of the modulator is given by%

\begin{equation}
T\left(  \mathcal{\varepsilon}\right)  =\left\vert \left\langle \psi_{p}%
|\psi_{f}\left(  \mathcal{\varepsilon}\right)  \right\rangle \right\vert ^{2}.
\end{equation}

Given the perturbed and unperturbed final states, $\left\vert \psi_{f}\left(
\mathcal{\varepsilon}\right)  \right\rangle $ and $\left\vert \psi_{f}\left(
0\right)  \right\rangle $ respectively, what is the optimum choice of a
normalized $\left\vert \psi_{p}\right\rangle $ that will maximize $\left\vert
dT/d\mathcal{\varepsilon}\right\vert $ ? \ Define the density operator%

\begin{equation}
\rho\left(  \mathcal{\varepsilon}\right)  =\left\vert \psi_{f}\left(
\mathcal{\varepsilon}\right)  \right\rangle \left\langle \psi_{f}\left(
\mathcal{\varepsilon}\right)  \right\vert ,
\end{equation}
and the operator%

\begin{equation}
\Delta\rho=\rho\left(  \mathcal{\varepsilon}\right)  -\rho\left(  0\right)  .
\end{equation}

For a small $\mathcal{\varepsilon}$ one has%

\begin{equation}
\frac{dT}{d\mathcal{\varepsilon}}=\frac{1}{\mathcal{\varepsilon}}\left\langle
\psi_{p}\right\vert \Delta\rho\left\vert \psi_{p}\right\rangle .
\end{equation}

The operator $\Delta\rho$ is Hermitian, thus $\left\vert \psi_{p}\right\rangle
$ that will maximize $\left\vert dT/d\lambda\right\vert $ is an eigenvector of
$\Delta\rho$ with the largest eigenvalue in absolute value. \ The eigenvalues
of $\Delta\rho$ are given by $\pm\sqrt{1-\left\vert \left\langle \psi
_{f}\left(  \mathcal{\varepsilon}\right)  |\psi_{f}\left(  0\right)
\right\rangle \right\vert ^{2}}$, thus%

\begin{equation}
\left\vert \frac{dT}{d\mathcal{\varepsilon}}\right\vert \leq\frac
{1}{\left\vert \mathcal{\varepsilon}\right\vert }\sqrt{1-\left\vert
\left\langle \psi_{i}\right\vert U^{\dag}\left(  0\right)  U\left(
\mathcal{\varepsilon}\right)  \left\vert \psi_{i}\right\rangle \right\vert
^{2}}.
\end{equation}
\qquad

Using perturbation expansion \cite{Weissbluth} one finds to 2nd order in
$\mathcal{\varepsilon}$%

\begin{align}
&  \left\langle \psi_{i}\right\vert U^{\dag}\left(  0\right)  U\left(
\mathcal{\varepsilon}\right)  \left\vert \psi_{i}\right\rangle \\
&  =1+i\mathcal{\varepsilon}%
{\textstyle\int\limits_{s_{0}}^{s_{1}}}
ds^{\prime}\left\langle \mathcal{K}_{1}\left(  s^{\prime}\right)
\right\rangle -\mathcal{\varepsilon}^{2}%
{\textstyle\int\limits_{s_{0}}^{s_{1}}}
ds^{\prime}%
{\textstyle\int\limits_{s_{0}}^{s^{\prime}}}
ds^{\prime\prime}\left\langle \mathcal{K}_{1}\left(  s^{\prime}\right)
\mathcal{K}_{1}\left(  s^{\prime\prime}\right)  \right\rangle ,\nonumber
\end{align}
where the symbol $\left\langle {}\right\rangle $ represents expectation value,
$\left\langle A\left(  s\right)  \right\rangle =\left\langle \psi
_{i}\right\vert A_{H}\left\vert \psi_{i}\right\rangle $ for a general operator
$A$, where $A_{H}$ is defined as%

\begin{equation}
A_{H}\left(  s\right)  \equiv u_{0}^{\dag}\left(  s,s_{0}\right)
Au_{0}\left(  s,s_{0}\right)
\end{equation}
and $u_{0}\left(  s,s_{0}\right)  $ is the $s$ evolution operator from $s_{0}$
to $s$ generated by $\mathcal{K}_{0}$. \ Since $\mathcal{K}_{1}\left(
s\right)  $ is Hermitian one finds to lowest order in $\mathcal{\varepsilon}$%

\begin{align}
&  \left\vert \left\langle \psi_{i}\right\vert U^{\dag}\left(  0\right)
U\left(  \mathcal{\varepsilon}\right)  \left\vert \psi_{i}\right\rangle
\right\vert ^{2}\label{nu(lambda)}\\
&  =1-\mathcal{\varepsilon}^{2}%
{\textstyle\int\limits_{s_{0}}^{s_{1}}}
ds^{\prime}%
{\textstyle\int\limits_{s_{0}}^{s_{1}}}
ds^{\prime\prime}\left\langle \Delta\mathcal{K}_{1}\left(  s^{\prime}\right)
\Delta\mathcal{K}_{1}\left(  s^{\prime\prime}\right)  \right\rangle ,\nonumber
\end{align}
where $\Delta\mathcal{K}_{1}\left(  s\right)  =\mathcal{K}_{1}\left(
s\right)  -\left\langle \mathcal{K}_{1}\left(  s\right)  \right\rangle $. Thus%

\begin{equation}
\left\vert \frac{dT}{d\mathcal{\varepsilon}}\right\vert ^{2}\leq\left\vert
{\textstyle\int\limits_{s_{0}}^{s_{1}}}
ds^{\prime}%
{\textstyle\int\limits_{s_{0}}^{s_{1}}}
ds^{\prime\prime}\left\langle \Delta\mathcal{K}_{1}\left(  s^{\prime}\right)
\Delta\mathcal{K}_{1}\left(  s^{\prime\prime}\right)  \right\rangle
\right\vert .
\end{equation}

This upper bound imposed upon the responsivity can be further simplified by
employing the Schwartz inequality%

\begin{equation}
\left\vert \frac{dT}{d\mathcal{\varepsilon}}\right\vert \leq%
{\textstyle\int\limits_{s_{0}}^{s_{1}}}
ds^{\prime}\sqrt{\left\langle \left[  \Delta\mathcal{K}_{1}\left(  s^{\prime
}\right)  \right]  ^{2}\right\rangle }. \label{dT/de}%
\end{equation}

\textit{Two-mode Case - }Consider the case where the dimensionality of
$\left\vert \psi\left(  s\right)  \right\rangle $ is two. \ Ignoring the
common phase factor, the Hermitian operator $\mathcal{K}_{1}$ can be assumed
to be traceless, thus it can be expressed as%

\begin{equation}
\mathcal{K}_{1}=\mathbf{\kappa}_{1}\cdot\mathbf{\sigma},
\end{equation}
where $\mathbf{\kappa}_{1}\mathbf{=}\left\vert \mathbf{\kappa}_{1}\right\vert
\mathbf{\hat{\kappa}}_{1}$ is a three-dimensional real vector with length
$\left\vert \mathbf{\kappa}_{1}\right\vert $ ($\mathbf{\hat{\kappa}}_{1}$ is a
unit vector) and the components of the Pauli matrix vector $\mathbf{\sigma}%
$\ \cite{Weissbluth} are given by:%

\begin{equation}
\sigma_{1}=\left(
\begin{array}
[c]{cc}%
0 & 1\\
1 & 0
\end{array}
\right)  ,\sigma_{2}=\left(
\begin{array}
[c]{cc}%
0 & -i\\
i & 0
\end{array}
\right)  ,\sigma_{3}=\left(
\begin{array}
[c]{cc}%
1 & 0\\
0 & -1
\end{array}
\right)  .
\end{equation}

It is straightforward to show that Eq. \ref{dT/de} for the present case yields%

\begin{equation}
\left\vert \frac{dT}{d\mathcal{\varepsilon}}\right\vert \leq%
{\textstyle\int\limits_{s_{0}}^{s_{1}}}
ds^{\prime}\left\vert \mathbf{\kappa}_{1}\left(  s^{\prime}\right)
\right\vert . \label{upper_bound}%
\end{equation}

Similar upper bound can be found for the angle $\theta$ between the
polarization unit vectors $\mathbf{p}\left(  \mathcal{\varepsilon}\right)
=\left\langle \psi_{i}\right\vert U^{\dag}\left(  \mathcal{\varepsilon
}\right)  \mathbf{\sigma}U\left(  \mathcal{\varepsilon}\right)  \left\vert
\psi_{i}\right\rangle $ and $\mathbf{p}\left(  0\right)  =\left\langle
\psi_{i}\right\vert U^{\dag}\left(  0\right)  \mathbf{\sigma}U\left(
0\right)  \left\vert \psi_{i}\right\rangle $ on the Bloch sphere. \ Using
\ref{nu(lambda)} and assuming the case $\theta<<1$ one finds%

\begin{equation}
\theta\leq2%
{\textstyle\int\limits_{s_{0}}^{s_{1}}}
ds^{\prime}\left\vert \mathcal{\varepsilon}\mathbf{\kappa}_{1}\left(
s^{\prime}\right)  \right\vert .
\end{equation}

Full modulation between $T=0$ and $T=1$ requires that the total change in
$\theta$ exceeds $\pi$ (assuming $\left\vert \psi_{p}\right\rangle $ is kept
fixed). \ Thus, if the applied birefringence strength is bounded by
$\left\vert \mathcal{\varepsilon}\mathbf{\kappa}_{1}\left(  s^{\prime}\right)
\right\vert \leq\kappa_{\max}$, full modulation occurs only for%

\begin{equation}
\Delta s\geq\frac{\pi}{2\kappa_{\max}}. \label{full_mod_up_bou}%
\end{equation}

\textit{Examples - }As a simple example, consider a modulator based on an
optical fiber. \ Circularly polarized light is injected into the fiber and a
polarizer located at the fiber end allows transmission of only linearly
polarized light. \ Modulation is achieved by applying linear birefringence
along some section of the fiber having length $\Delta s$.

For the present example we chose $\left\vert \psi\right\rangle =\left\vert
+;\hat{2}\right\rangle $ at $s=0$ ($\left\vert \pm;\mathbf{\hat{u}%
}\right\rangle $, with $\mathbf{\hat{u}}$ being a unit vector, denotes an
eigenvector of $\mathbf{\sigma}\cdot\mathbf{\hat{u}}$ with an eigenvalue
$\pm1$), and the polarizer state is $\left\vert \psi_{p}\right\rangle
=\left\vert +;\hat{3}\right\rangle $. \ Moreover, $\mathcal{K}_{0}=0$ and
$\mathcal{K}_{1}=\mathbf{\kappa}\cdot\mathbf{\sigma}$, where $\mathbf{\kappa
=}\left(  1/2\right)  \left(  k_{1},0,0\right)  $. \ Integrating the equation
of motion yields%

\begin{equation}
T\left(  \varepsilon\right)  =\sin^{2}\left(  \frac{\varepsilon k_{1}s_{1}}%
{2}-\frac{\pi}{4}\right)  .
\end{equation}

Thus at $\varepsilon=0$ the derivative $\left\vert dT/d\varepsilon\right\vert
$ approaches the upper bound given in \ref{upper_bound}. \ Moreover, full
modulation is obtained for $s_{0}=-\pi/2\varepsilon k_{1}$, and $s_{1}%
=\pi/2\varepsilon k_{1}$, thus also for this case the upper bound given by
\ref{full_mod_up_bou} is achieved.

The next example deals with a modulator based on a transition between
adiabatic to non-adiabatic regimes, as in Ref. \cite{Buks}. \ Consider the
case where $\mathcal{K}=\mathbf{\kappa}\cdot\mathbf{\sigma}$,%

\begin{equation}
\mathbf{\kappa}\left(  s\right)  =\gamma\left(  0,\sqrt{\lambda^{2}-\left(
\gamma s\right)  ^{2}},\gamma s\right)  ,
\end{equation}
where $\gamma$ is a real constant with dimensionality of 1/length, $\lambda$
is a non-negative dimensionless real parameter, and $\left\vert \gamma
s\right\vert \leq\lambda$.

Consider the case where for $s_{0}=-\lambda/\gamma$ the state of the system is
a local eigenstate of $\mathcal{K}\left(  s\right)  $ with positive
eigenvalue, namely $\left\vert \psi\left(  s_{0}\right)  \right\rangle
=\left\vert -;\hat{3}\right\rangle $. \ When $\lambda>>1$ the state evolves
adiabatically \cite{Berry} and remains parallel to $\mathbf{\kappa}\left(
s\right)  $. \ The polarizer is located at $s_{1}=\lambda/\gamma$ and its
state is given by $\left\vert \psi_{p}\right\rangle =\left\vert -;\hat
{3}\right\rangle $. \ Thus in the adiabatic limit $T=0$. \ Approximation
solution for the case $\lambda\gtrsim1$ can be found by considering the lowest
order correction to the adiabatic limit \cite{Migdal}, \cite{Buks}. \ The
value of $T$ is the probability of Zener transition to occur which can be
calculated to lowest order%

\begin{equation}
T\simeq\frac{\pi^{2}}{4}J_{0}^{2}\left(  2\lambda^{2}\right)  \text{
\ \ \ \ }\left(  \text{for \ }\lambda\gtrsim1\right)  .
\label{T vs lambda adiabatic}%
\end{equation}

\begin{figure}
[ptb]
\begin{center}
\includegraphics[
height=1.5592in,
width=3.1272in
]%
{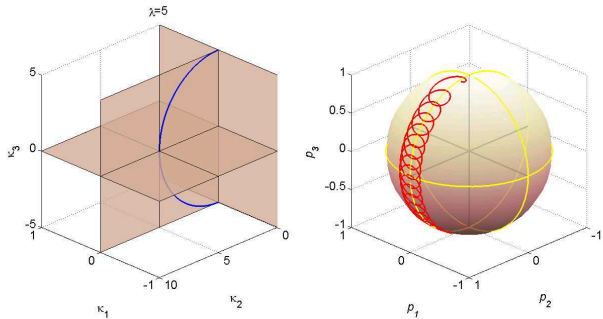}%
\caption{Example of numerical integration of the equation of motion for the
case $\lambda=5$. \ On the left the curve $\mathbf{\kappa}\left(  s\right)  $
is shown \ and on the right the evolution of the polarization vector
$\mathbf{p}\left(  s\right)  $ on the Bloch sphere is depicted.}%
\label{Numercal integration}%
\end{center}
\end{figure}

\begin{figure}
[ptb]
\begin{center}
\includegraphics[
height=2.444in,
width=3.0424in
]%
{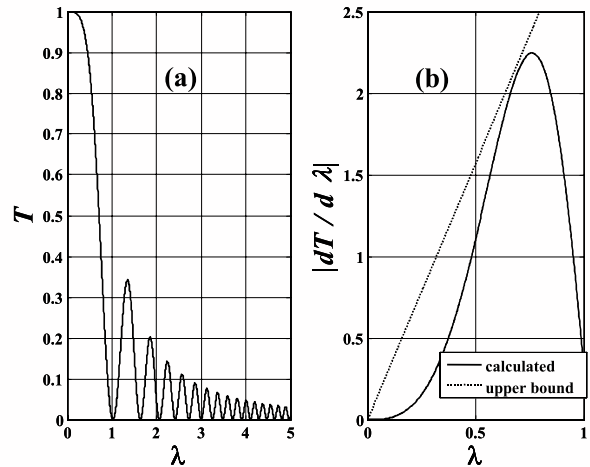}%
\caption{Calculated and upper bound of responsivity. \ (a) Numerical
calculation of $T$ vs. $\lambda$. \ (b) Comparison between the calculater
$\left\vert dT/d\lambda\right\vert $ and the upper bound given by Eq.
\ref{dT/d lamnda UB}.}%
\label{T vs lambda}%
\end{center}
\end{figure}

This approximation is compared with the calculated value of $T\left(
\lambda\right)  $ obtained from numerical integration of the equation of
motion. \ The case $\lambda=5$ is presented in Fig. \ref{Numercal integration}
as an example. \ Figure \ref{T vs lambda} (a) shows the calculated value of
$T\left(  \lambda\right)  $ in the range $0\leq\lambda\leq5$. \ Comparing the
approximated result in Eq. \ref{T vs lambda adiabatic} with the numerical
solution shows, as expected, good agreement for $\lambda\gtrsim1$.

On the other hand, the upper bound given by Eq. \ref{upper_bound} for this
case reads%

\begin{equation}
\left\vert \frac{dT}{d\lambda}\right\vert \leq%
{\textstyle\int\limits_{-\lambda_{0}/\gamma}^{\lambda_{0}/\gamma}}
ds^{\prime}\frac{\gamma\lambda_{0}}{\sqrt{\lambda_{0}^{2}-\left(  \gamma
s^{\prime}\right)  ^{2}}}=\pi\lambda_{0}. \label{dT/d lamnda UB}%
\end{equation}

Comparison between the numerically calculated $\left\vert dT/d\lambda
\right\vert $ and the above upper bound is seen in Fig. \ref{T vs lambda} (b).
\ Contrary to the previous example, in this case the upper bound is not
reached for any value of $\lambda$. \ However, in the transition region,
between the adiabatic and non-adiabatic limits, near $\lambda=0.695$ the
responsivity is only some 2\% below the upper bound. \ Similarly, for the
modulator discussed in Ref. \cite{Buks}, it was found that largest
responsivity is obtained in the transition region between adiabatic and
non-adiabatic limits.

Note that the bounds discussed in this paper can be employed for other linear
systems. \ For example, the same analysis may lead to a lower bound imposed
upon the time required for performing a given quantum gate on a system of
quantum bits in a quantum computer, when the maximum perturbation strength is given.%

%

The author thanks Avishai Eyal for very useful and stimulating discussions.

\newpage
\bibliographystyle{plain}
\bibliography{apssamp}

\end{document}